\pdfoutput=1
\documentclass[twocolumn,preprintnumbers,superscriptaddress,aps]{revtex4-1}

\usepackage{amsthm,amsmath,amssymb,physics,mathtools}
\usepackage{bbm}

\usepackage{array,esint}
\def\CMD#1{%
   $ \csname#1\endcsname \displaystyle\csname#1\endcsname $ & \texttt{\textbackslash#1} &}

\usepackage{hyperref}
\hypersetup{
    colorlinks = true,       % false: boxed links; true: colored links
    linkcolor = red,          % color of internal links (change box color with linkbordercolor)
    citecolor = magenta,        % color of links to bibliography
    filecolor=red,      % color of file links
    urlcolor= cyan           % color of external links
}

\usepackage{graphicx}
\graphicspath{ {TBA-PRL/} }
\usepackage{xcolor}

\usepackage{dsfont}

\newcommand{\be}{\begin{equation}}
\newcommand{\ee}{\end{equation}}
\newcommand{\beq}{\begin{equation}}
\newcommand{\eeq}{\end{equation}}
\newcommand{\bea}{\begin{equation} \begin{aligned}}
\newcommand{\eea}{\end{aligned} \end{equation}}

\newcommand{\1}{1}
\newcommand{\K}{\mathbb{K}}
\newcommand{\R}{\mathcal{R}}
\newcommand{\Gc}{\Gamma_{\textrm{cusp}}}
\newcommand{\Go}{\Gamma_{\textrm{oct}}}
\newcommand{\Gt}{\Gamma_{\textrm{hex}}}
\newcommand{\W}{\mathcal{W}}
\newcommand{\N}{\mathcal{N}}

\newcommand{\Q}{\mathbb{Q}}

\newcommand{\F}{F}
\newcommand{\G}{M}
\newcommand{\E}{\mathcal{E}}

\newcommand{\ca}{c}
\newcommand{\sa}{s}

\begin{document}

\preprint{SLAC--PUB--17501}
\preprint{DESY 20-002}

\title{The Origin of the Six-Gluon Amplitude in Planar $\mathcal{N}=4$ SYM}
\author{Benjamin Basso}
\email{benjamin.basso@phys.ens.fr}
\affiliation{Laboratoire de physique de l'Ecole normale sup\'erieure, ENS, Universit\'e PSL, CNRS, Sorbonne Universit\'e, Universit\'e Paris-Diderot, Sorbonne Paris Cit\'e, 24 rue Lhomond, 75005 Paris, France}
\author{Lance J. Dixon}
\email{lance@slac.stanford.edu}
\affiliation{SLAC National Accelerator Laboratory, Stanford University, Stanford, CA 94309, USA}
\author{Georgios Papathanasiou}
\email{georgios.papathanasiou@desy.de}
\affiliation{DESY Theory Group, DESY Hamburg, Notkestra{\ss}e 85, D-22607 Hamburg, Germany}

\date{\today}

\begin{abstract}
We study the maximally-helicity-violating (MHV) six-gluon scattering amplitude
in planar $\mathcal{N} = 4$ super-Yang-Mills theory at finite coupling
when all three cross ratios are small. It exhibits a double logarithmic
scaling in the cross ratios, controlled by a handful of ``anomalous dimensions"
that are functions of the coupling constant alone.
Inspired by known seven-loop results at weak coupling and the
integrability-based pentagon OPE, we present conjectures for the
all-order resummation of these anomalous dimensions.  At strong coupling,
our predictions agree perfectly with the string theory analysis. Intriguingly,
the simplest of these anomalous dimensions coincides with one
describing the light-like limit of the octagon, namely the four-point function of
large-charge BPS operators.
\end{abstract}

\maketitle

\section{\label{sec:intro}Introduction}

\begin{center}
\begin{table*}
\begin{tabular}{|c|c|c|c|c|c|}
  \hline
  & $L= 1$ & $L = 2$ & $L = 3$ & $L = 4$ & $L = 5$ \\
  \hline
  $\Go$ & 4 & $- 16\zeta_{2}$ & $256 \zeta_{4}$ & $-3264 \zeta_{6}$ & $\frac{126976}{3} \zeta_{8}$ \\
  $\Gc$ & 4 & $- 8\zeta_{2}$ &  $88 \zeta_{4}$ &  $-876\zeta_{6} - 32 \zeta_3^2$ & $\frac{28384}{3} \zeta_{8} + 128\zeta_{2}\zeta_{3}^2 + 640 \zeta_3 \zeta_5$ \\
  $\Gt$ & 4 & $- 4\zeta_{2}$ & $ 34 \zeta_{4}$ & $ -\frac{603}{2}\zeta_{6} - 24 \zeta_3^2$ & $\frac{18287}{6} \zeta_{8} + 48\zeta_{2} \zeta_3^2 + 480 \zeta_3 \zeta_5$ \\
  $C_0$ & $- 3\zeta_{2}$ & $\frac{77}{4} \zeta_{4}$ & $ -\frac{4463}{24} \zeta_{6} + 2\zeta_{3}^2$ & $ \frac{67645}{32} \zeta_{8} + 6\zeta_{2}\zeta_{3}^2 -40\zeta_{3} \zeta_{5}$ & $-\frac{4184281}{160}\zeta_{10}-65\zeta_{4}\zeta_{3}^2-120 \zeta_2 \zeta_3 \zeta_5+228 \zeta_{5}^2+420\zeta_{3}\zeta_{7}$ \\
  \hline
\end{tabular}
\caption{\label{tab:loops} Coefficients of expansions in $g^2$ of
  the main coefficients through $L = 5$ loops.}
\end{table*}
\end{center}

The scattering of massless gluons in maximally supersymmetric
gauge theory, $\mathcal{N} = 4$ super-Yang-Mills theory (SYM),
exhibits remarkable simplifications in the planar limit of
a large number of colors.  Scattering amplitudes for $n$ gluons become dual
to null polygonal Wilson loops~\cite{Alday:2007hr,Drummond:2007aua,
  Brandhuber:2007yx,Drummond:2007cf,Drummond:2007au} and consequently
they depend essentially only on $3n-15$ dual conformal cross
ratios~\cite{Bern:2008ap,Drummond:2008aq},
out of the $3n-10$ Mandelstam invariants.
Powerful bootstrap techniques~\cite{Dixon:2011pw,Dixon:2013eka,
Dixon:2014iba,Dixon:2015iva,Caron-Huot:2016owq}
allow the construction of the six-gluon maximally-helicity-violating
(MHV) amplitude through seven loops,
and the next-to-MHV amplitude through six loops~\cite{Caron-Huot:2019vjl}.
Seven-point amplitudes have also been bootstrapped through four
loops~\cite{Drummond:2014ffa,Dixon:2016nkn,Drummond:2018caf}
at the level of the symbol~\cite{Goncharov:2010jf}.

For generic values of the cross ratios, the perturbative results
can be expressed in terms of generalized polylogarithms to all orders,
but resumming the results into a finite-coupling expression remains
challenging.  In the near-collinear limit, a finite-coupling description
is available, based on integrability and the pentagon operator product
expansion (OPE)~\cite{Alday:2010ku,Basso:2013vsa,Basso:2013aha,Basso:2014koa,Basso:2014nra,Belitsky:2014sla,Belitsky:2014lta,Basso:2015uxa}.

In this letter we will provide a (conjectural) finite-coupling description
for another kinematical limit of the six-gluon MHV amplitude,
where all three cross ratios become small. The ``origin" is reached,
roughly speaking, by taking three adjacent pairs of gluon momenta to be
parallel (collinear) simultaneously. However,
it is a Euclidean limit, which cannot be achieved for real Minkowski momenta.
Our description of the amplitude at the origin
is based on resumming the OPE for a gas of
gluonic flux-tube excitations. It involves a ``tilted'' version
of the Beisert-Eden-Staudacher (BES) kernel entering
the finite-coupling formula for the cusp anomalous
dimension~\cite{Beisert:2006ez}.  Different tilt angles generate
different anomalous dimensions controlling logarithmically-enhanced
terms in the amplitude.
Intriguingly, one of the anomalous dimensions
also appears in the light-like limit of the
octagon~\cite{Coronado:2018ypq,Coronado:2018cxj,Kostov:2019stn,
  Kostov:2019auq,Belitsky:2019fan,Bargheer:2019exp},
a correlation function of four operators with large $R$ charge.
We also predict the non-logarithmic term, as well as the coefficient $\rho$
controlling a ``cosmic'' amplitude normalization~\cite{Caron-Huot:2019bsq}.
Our key results are eqs.~(\ref{eq:Kalpha})--(\ref{eq:Dalpha}) for the anomalous
dimensions and constant terms, and eq.~(\ref{eq:norm}) for ${\cal N}=\rho$.

More precisely, we consider the MHV amplitude normalized by the BDS-like
ansatz~\cite{Bern:2005iz,Alday:2009dv,Dixon:2015iva,Caron-Huot:2016owq},
which remains finite as the dimensional regulator
$\epsilon = 2-\tfrac{1}{2}D \rightarrow 0$,
\beq
\label{eq:EtoR}
\mathcal{E}(u_i)
= \lim_{\epsilon \rightarrow 0} \frac{\mathcal{A}_{6} (s_{ij}, \epsilon)}
{\mathcal{A}_{6}^{\textrm{{\tiny BDS-like}}} (s_{ij}, \epsilon)}
=  \exp[\R_{6} + \tfrac{1}{4}\Gc \E^{(1)}] \,.
\eeq
The notation and normalization (for now) follow
ref.~\cite{Caron-Huot:2016owq},
where $\Gc$ is the cusp anomalous dimension,
$\R_{6}$ is the remainder function,
and $\E^{(1)} = \sum_{i=1}^{3}\textrm{Li}_{2}(1-1/u_{i})$ is the one-loop
amplitude with $\textrm{Li}_{2}$ the dilogarithm.  The normalized
amplitude is a function of three cross ratios,
\beq
u_1 = \frac{s_{12} s_{45}}{s_{123} s_{345}}\,,
\,\, u_{2} = \frac{s_{23}s_{56}}{s_{234}s_{123}}\,,
\,\, u_{3} = \frac{s_{34}s_{61}}{s_{345}s_{234}}\, , 
\eeq
constructed from the Mandelstam invariants $s_{i\ldots j} = (p_i+\ldots +p_j)^2$.

The logarithm of the amplitude $\E$,
or equivalently the remainder function $\R_6$,
exhibits logarithmic scaling when all cross ratios $\rightarrow 0$,
\beq\label{eq:main}
\begin{aligned}
\ln{\E} &= -\frac{\Go}{24}\ln^{2}{(u_1 u_2 u_3)} -\frac{\Gt}{24} \sum_{i=1}^{3} \ln^{2}{(\frac{u_i}{u_{i+1}})} \\
&\hskip0.5cm
+ C_{0} + {\cal O}(u_i),
\end{aligned}
\eeq
with $u_4 \equiv u_1$ and where $\Go, \Gt$ and $C_0$ are functions of the
coupling constant $g^2 = \lambda/(4\pi)^2$ of the planar theory.
The simpler $\ln^2u$ behavior on the diagonal $u_1 = u_2 = u_3 = u$,
where $\Gt$ drops out,
was conjectured \cite{Alday:2009dv} to hold at any coupling for a function
$h = -\tfrac{3}{8}(\Go-\Gc)$ appearing in $\R_6$,
based on two-loop results in gauge theory and strong coupling behavior
in string theory. The more general behavior (\ref{eq:main})
for unequal $u_i$ was observed through seven loops~\cite{Caron-Huot:2019vjl},
up to power corrections in the $u_i$.
Its structure is reminiscent of Sudakov double-logarithms.

The subleading power corrections to eq.~(\ref{eq:main}) do not exponentiate
simply; in $\ln{\E}$ at $L$ loops at finite $u_1$ there are terms with
up to $L$ powers of $\{ \ln u_2, \ln u_3 \}$.  From this observation,
based on results in ref.~\cite{Caron-Huot:2019vjl}, we expect
the simplest finite-coupling resummation, apart from OPE limits, to be
when all three cross ratios are small.

%%%%%%%%%%%%%%%%%%%%%%%%%%%%%%%%
\section{\label{sec:data} Weak coupling evidence}

The first evidence for eq.~(\ref{eq:main}) comes from weak coupling.
The hexagon function bootstrap enables the analytic determination
of $\R_6$ through seven
loops~\cite{Dixon:2011pw,Dixon:2013eka,Dixon:2014voa,Caron-Huot:2016owq,
  Caron-Huot:2019vjl},
throughout the entire kinematical space.  At the origin, the remainder
function admits a simple representation, through at least
seven loops \cite{Caron-Huot:2019vjl},
\beq
\R_{6} = c_{1} P_{1} + c_{2} P_{2} + c_0 + {\cal O}(u_i) \,,
\eeq
in terms of the two symmetric quadratic polynomials in $\ln{u_i}$,
\beq
P_{1} = P_{2} + \sum_{i=1}^{3} \ln^2{u_{i}}\, ,
\quad P_{2} = \sum_{i=1}^3 \ln{u_{i}}\ln{u_{i+1}}\,.
\eeq
There is no term linear in $\ln{u_i}$.
Close to the origin,
$\mathcal{E}^{(1)} = -\frac{1}{2}\sum_{i}\ln^2{u_{i}}-3\zeta_{2}$,
and using eq.~(\ref{eq:EtoR}), one finds
\beq
\Go = \Gc - 16 c_{1} - 8c_{2} \, , \quad \Gt = \Gc - 4c_{1} + 4c_{2} \, ,
\eeq
and $C_0 = c_0-\frac{3}{4}\zeta_2\Gc$. Perturbative results in
Section 4.2 of ref.~\cite{Caron-Huot:2019vjl} yield the numbers in
table \ref{tab:loops} for the expansion in $g^2$, truncated here
to 5 loops due to space limitations, where $\zeta_{n} = \zeta(n)$
is the Riemann zeta function. Note that $\Go$ has an expansion in powers
of $\pi^2$ only (through 7 loops at least). Furthermore, it agrees
with the exact~\cite{Belitsky:2019fan} anomalous dimension controlling
the light-like limit of the
octagon~\cite{Coronado:2018ypq,Coronado:2018cxj,Kostov:2019stn,Kostov:2019auq},
\beq
\label{eq:exact}
\Go = \frac{2}{\pi^2}\ln{\cosh{(2\pi g)}}\, .
\eeq
The other quantities are more complicated. Their perturbative expansions contain products of odd Riemann zeta values, much like the cusp anomalous dimension, which is recalled in the table.

%%%%%%%%%%%%%%%%%%%%%%%%%%%%%%%%
\section{\label{sec:fluxtube} Pentagon OPE}

\begin{figure}
\begin{center}
\includegraphics[scale=0.25]{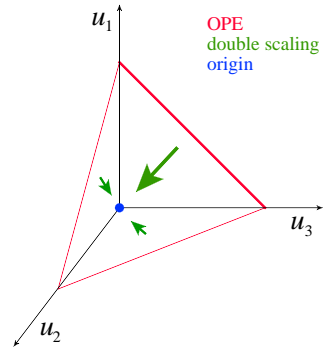}
\end{center}
\vspace{-2cm}
\caption{Six-gluon kinematics. The collinear OPE is an expansion around one edge of the triangle, e.g.~around $u_2 = 0$ and $u_1+u_3 = 1$. The latter condition must be relaxed to get to the origin, as discussed below eq.~(\ref{eq:tau}).}\label{Corner} 
\end{figure}

Insight at higher loops is provided by the pentagon
OPE~\cite{Basso:2013vsa}. It generates a systematic expansion of the
amplitude around the collinear limit, $u_2\to0$, $u_1+u_3\to1$,
see fig.~\ref{Corner}, based on (flux tube) excitations of
the dual two-dimensional string theory
of 't Hooft surfaces that emerges in the large $N_c$, planar limit.
The collinear limit is $\tau\to\infty$
at fixed $\sigma$ and $\varphi$ with the parametrization
\beq\label{eq:tau}
\begin{aligned}
&u_{2} = \frac{1}{e^{2\tau}+1}\, , \qquad u_{1}  =  e^{2\tau+2\sigma}u_{2}u_{3}\, , \\
&u_{3} = \frac{1}{1+e^{2\sigma}+2e^{\sigma-\tau}\cosh{\varphi}+e^{-2\tau}}\, .
\end{aligned}
\eeq
Because $\tau$ is conjugate to the energy, or twist, of flux-tube excitations,
the collinear limit is controlled by the lowest-twist excitations
that can propagate from one side of the hexagon to the other.
These include gluonic, fermionic and scalar excitations.
Higher-twist contributions are suppressed by additional powers of $e^{-\tau}$.

We can move from the collinear limit toward the origin by first considering the limit where $\varphi, \tau$ are taken to be large, keeping their difference finite~\cite{Basso:2014nra,Drummond:2015jea}. In this double-scaling limit, $u_2\to0$, but $u_1$ and $u_3$ are generic. The hyperbolic angle $\varphi$ is conjugate to the helicity of the particles exchanged
in the OPE channel. As $\varphi \to \infty$, the OPE is dominated by
gluonic excitations, which have the highest helicity for a given twist.
They form a family labelled by an integer $a=1, 2, \ldots$, and each
carries a rapidity $u$ for its energy $E_a(u)$ and momentum $p_a(u)$ (conjugate to $\sigma$).

The OPE is naturally expressed in terms of the framed Wilson-loop
expectation value $\W_6$~\cite{Alday:2010ku}, which is related to $\E$ by
\beq\label{eq:framedW}
\W_6 = \E \, \exp[\frac{1}{2}\Gc (\sigma^2+ \tau^2+\zeta_2) ] \, .
\eeq
In the double-scaling limit, where only gluonic excitations contribute, $\W_6$ takes the form,
\beq\label{eq:OPE}
\W_{6} = \sum_{N=0}^{\infty}\frac{1}{N!} \sum_{\textbf{a}} e^{\varphi\sum_{k=1}^{N}a_k} \int \frac{d\textbf{u}}{(2\pi)^N} \frac{e^{-\tau E+i\sigma P}\prod_k \mu_k}{\prod_{k<l}P_{kl}P_{lk}}
\eeq
where $\textbf{a} = (a_1, \ldots, a_N)$ are positive integers and $d\textbf{u} = du_1 \ldots du_N$ with $u_k\in \mathbb{R}$. The total energy and momentum of the $N$-gluon flux-tube state are $E = \sum_{k}E_{a_k}(u_k)$ and $P = \sum_{k}p_{a_k}(u_k)$. The integrand is built out of the pentagon transitions $P_{kl} = P_{a_{k}|a_{l}}(u_{k}|u_{l})$ and measures $\mu_k = \mu_{a_{k}}(u_k)$, which have been conjectured to all orders in the coupling constant~\cite{Basso:2014nra}. This concludes our review of the double-scaling limit.

To get to the origin from the double-scaling limit, we must then take $\varphi-\tau \rightarrow \infty$. While this limit lies outside of the radius of convergence of the OPE series (\ref{eq:OPE}), we may nevertheless reach it by analytically continuing in the helicity $a$, and replacing the sum by a contour integral with the help of the Sommerfeld-Watson transform,
\beq\label{eq:Sommerfeld-Watson}
\sum_{a \geqslant 1} (-1)^{a} f(a) \rightarrow \int\limits_{\epsilon -i\infty}^{\epsilon +i\infty} \frac{if(a)da}{2\sin{(\pi a)}}\, ,
\eeq
with $\epsilon \in (0, 1)$ \footnote{More precisely, the initial integration contour goes from slightly below to slightly above the positive real axis. Along the lines of \cite{Drummond:2015jea}, it can be shown that $f(a)$ is a meromorphic function of $a$ to all orders in perturbation theory, and that its poles lie on the real axis. We may thus deform the contour as shown in eq.~\eqref{eq:Sommerfeld-Watson}.}.  Next we deform the contour of the $a$-integral to the left, picking up residues from poles with ${\rm Re}(a)\leq0$.  Because of the factor $e^{\varphi\sum_{k=1}^N a_k}$ in eq.~(\ref{eq:OPE}), poles with ${\rm Re}(a)<0$ are suppressed by powers of the $u_i$ near the origin at weak coupling.  That is, computing the $a = 0$ residue alone suffices to obtain the logarithmic and constant terms at the origin.

Take for illustration the one-loop $N=1$ result \cite{Alday:2010ku,Gaiotto:2011dt},
\beq
\mu_{a}(u) = (-1)^{a}\frac{g^2\Gamma(\tfrac{a}{2}+ i u)\Gamma(\tfrac{a}{2}- i u)}{(\frac{a^2}{4}+u^2)\Gamma(a)} + O(g^4)
\eeq
with $E_{a} = a + O(g^2)$ and $p_{a} = 2u +O(g^2)$. This integrand vanishes at $a=0$. Nonetheless, the $u$-integral diverges as $1/a^2$ due to pinch singularities at $u = \pm ia/2$.
Accordingly, the dominant contribution is obtained by considering the residue around either one of these singularities, say the one at $u = ia/2$. Doing the $u$-integral around $ia/2$ and then the $a$-integral around $0$, we get
\beq
\begin{aligned}
&i\varointctrclockwise\frac{dadu}{(2\pi)^2} e^{a\varphi-a\tau+2i u \sigma}\frac{\Gamma(1-a)\Gamma(\tfrac{a}{2}+i u)\Gamma(\tfrac{a}{2}-i u)}{\frac{a^2}{4}+u^2} \\ 
&\,\, = \sigma^2-(\varphi-\tau)^2-\zeta_{2} = -\ln{u_{1}}\ln{u_{3}}-\zeta_{2}\,,
\end{aligned}
\label{toyintegral}
\eeq
in agreement with the one-loop result $\E^{(1)}+2(\sigma^2+\tau^2+\zeta_{2})$ close to the origin, $u_i \rightarrow 0$, where we have
\beq\label{u123atorigin}
u_{1} \sim e^{\tau - \varphi + \sigma}, \quad
u_{2}\sim e^{-2\tau}, \quad
u_{3} \sim e^{\tau - \varphi - \sigma}.
\eeq
The above analysis remains unchanged as we increase the loop order or particle number: The amplitude at the origin may be obtained to all loops as the contour integral of the OPE integrand first around $u_k=ia_k/2$, and then around $a_k=0$, for $k=1,\ldots,N$. Since $N$-particle states are suppressed as $g^{2N^2}$, by restricting to $N\le 2$ and applying the techniques of \cite{Papathanasiou:2013uoa,Papathanasiou:2014yva,Drummond:2015jea} we indeed reproduce all existing data, and obtain new predictions at 8 loops.

At finite coupling, the pole at $u = ia/2$ is replaced by a square-root branch cut between $\pm 2g +i a/2$, and the recipe is to integrate $u$ closely around this cut. Equivalently, we may bring the contour through the cut to the so-called Goldstone sheet \cite{Basso:2011rc,Basso:2014nra}, where the flux-tube ingredients greatly simplify. 

In particular, all $\Gamma$ functions in the integrand \eqref{eq:OPE} disappear when passing to the Goldstone sheet, which in turn allows us to reexpress $\mathcal{W}_6$ as a simpler, infinite-dimensional integral. Strikingly, this integral is secretly Gaussian in the vicinity of the origin (but not away from it). As a result, it is entirely characterized by a small number of moments, for which we were able to obtain conjectures that matched the structure through four loops. The details of this technical analysis are relegated to Appendix~\ref{appx:Gaussian}.  After some elementary algebra, we can recast our conjectures for the moments as very concise expressions for the anomalous dimensions and constants appearing at the origin, to be described next. They feature the celebrated BES kernel~\cite{Beisert:2006ez} which enters the all-loop formula for the cusp anomalous dimension and is ubiquitous in the flux-tube dynamics.

%%%%%%%%%%%%%%%%%%%%%%%%%%
\section{Tilted BES kernel}\label{sec:conjectures}

The BES kernel can be described~\cite{Beisert:2006ez,Benna:2006nd,Alday:2007qf,Basso:2013aha}, after an
expansion in terms of Bessel functions, as a semi-infinite matrix,
\beq\label{eq:K}
\begin{aligned}
&\K_{ij} = 2j(-1)^{ij+j} \int\limits_{0}^{\infty}\frac{dt}{t}\frac{J_{i}(2gt)J_{j}(2gt)}{e^{t}-1}\, , \\
\end{aligned}
\eeq
where $J_{i}(z)$ is the $i$-th Bessel function of the first kind.
We can spell out our finite-coupling conjectures for the origin in terms of
this matrix.  To this end, let us partition $\K$ into four blocks according
to whether $i$ and $j$ in eq.~(\ref{eq:K}) are odd or even.
After reshuffling lines and columns, we write
\beq\label{eq:blocks}
\K = \left[\begin{array}{cc} \K_{\circ\circ} & \K_{\circ \bullet} \\ \K_{\bullet \circ} & \K_{\bullet\bullet} \end{array}\right]\, ,
\eeq
with $\K_{\circ\circ}$ the odd-odd block,
built out of overlaps of odd Bessel functions ($J_{2i-1}$),
$\K_{\circ\bullet}$ the odd-even one, and so on.

The tilted kernel is defined by
\beq\label{eq:Kalpha}
\K(\alpha) = 2\cos{\alpha} \left[\begin{array}{cc} \cos{\alpha}\, \K_{\circ\circ} & \sin{\alpha}\, \K_{\circ\bullet} \\ \sin{\alpha}\, \K_{\bullet \circ} & \cos{\alpha}\, \K_{\bullet\bullet} \end{array}\right]\, .
\eeq
It reduces to the BES kernel (\ref{eq:blocks}) when $\alpha = \pi/4$, that is $\K = \K(\pi/4)$.  Our conjectures are that the coefficients in (\ref{eq:main}) are given by
\beq\label{eq:G-alpha}
\Gamma_{\alpha} = 4g^2 \left[\frac{1}{1+\K(\alpha)}\right]_{11}
\eeq
with $\alpha = 0, \pi/4$ and $\pi/3$ for $\Go, \Gc$ and $\Gt$, respectively, where the ``11'' subscript denotes the top left component of the semi-infinite matrix.

The constant $C_0$ is more complicated as it arises from determinants of quadratic forms appearing in the secretly Gaussian integral. Using formulae for the determinants of block matrices, we get
\beq\label{eq:C0}
C_0 = -\frac{\zeta_{2}}{2}\Gc + D(\pi/4)-D(\pi/3)-\frac{1}{2}D(0) \, ,
\eeq
where
\beq\label{eq:Dalpha}
D(\alpha) \equiv \ln{\textrm{det}\, [1+\K(\alpha)]} = \textrm{tr}\, \ln{[1+\mathbb{K}(\alpha)]}\, .
\eeq

These formulae can be verified easily at weak coupling, since the matrix elements $\K_{ij} = O(g^{i+j})$. (See e.g.~Appendix A.2 in Ref.~\cite{Basso:2014nra} for explicit expressions.)  The inversion in (\ref{eq:G-alpha}) is done by expanding the geometric series in $\K(\alpha)$.  Through four loops we get
\begin{align}
\frac{\Gamma_{\alpha}}{4g^2} &= 1 -4 \ca^2 \zeta_{2}  g^2 + 8\ca^2(3+5\ca^2)\zeta_{4} g^4 \label{eq:Ga}\\
& -8\ca^2\left[(25 + 42\ca^2+35\ca^4)\zeta_{6}  + 4\sa^2\, \zeta_{3}^2\right]g^6 +\ldots\, ,\nonumber\\
D(\alpha) &= 4\ca^{2} \zeta_{2} g^{2} -4\ca^2(3+5\ca^2)\zeta_{4} g^{4}\label{eq:Da} \\
& + \frac{8}{3}\ca^2\left[(30 + 63\ca^2+35\ca^4)\zeta_{6}  + 12\sa^2\, \zeta_{3}^2\right]g^6 +\ldots \nonumber\, ,
\end{align}
where $\ca = \cos{\alpha}$, $\sa = \sin{\alpha}$, and we verify agreement with the numbers in table \ref{tab:loops} using eq.~(\ref{eq:C0}). Higher loops are easily generated.  We provide results through 25 loops in an ancillary file.  From the growth rate of their perturbative coefficients, all these quantities appear to have same radius of convergence, $g^2_{c} = 1/16$, as $\Gc$ \cite{Beisert:2006ez}.

The point $\alpha=0$ corresponds to the octagon \cite{Coronado:2018ypq,Coronado:2018cxj,Kostov:2019stn,Kostov:2019auq}. Here the off-diagonal blocks of the BES kernel drop out,
\beq
\mathbb{K}(\alpha = 0) = \left[\begin{array}{cc} 2\K_{\circ\circ} & 0 \\ 0 & 2\K_{\bullet\bullet}\end{array}\right]\, ,
\eeq
and with them all zeta values with odd arguments, leaving only powers of $\pi^2$. Nicely, in eq.~(\ref{eq:G-alpha}) these can be resummed
exactly \cite{Belitsky:2019fan} into eq.~(\ref{eq:exact}) for $\Go$,
and similarly for the associated determinant,
\beq
\label{eq:exactD}
D(0) = \frac{1}{4}\ln{\left[\frac{\sinh{(4\pi g)}}{4\pi g}\right]}\,,
\eeq
which also appears in the light-like octagon~\cite{Belitsky:2019fan}.

\begin{figure}
\begin{center}
\includegraphics[scale=0.45]{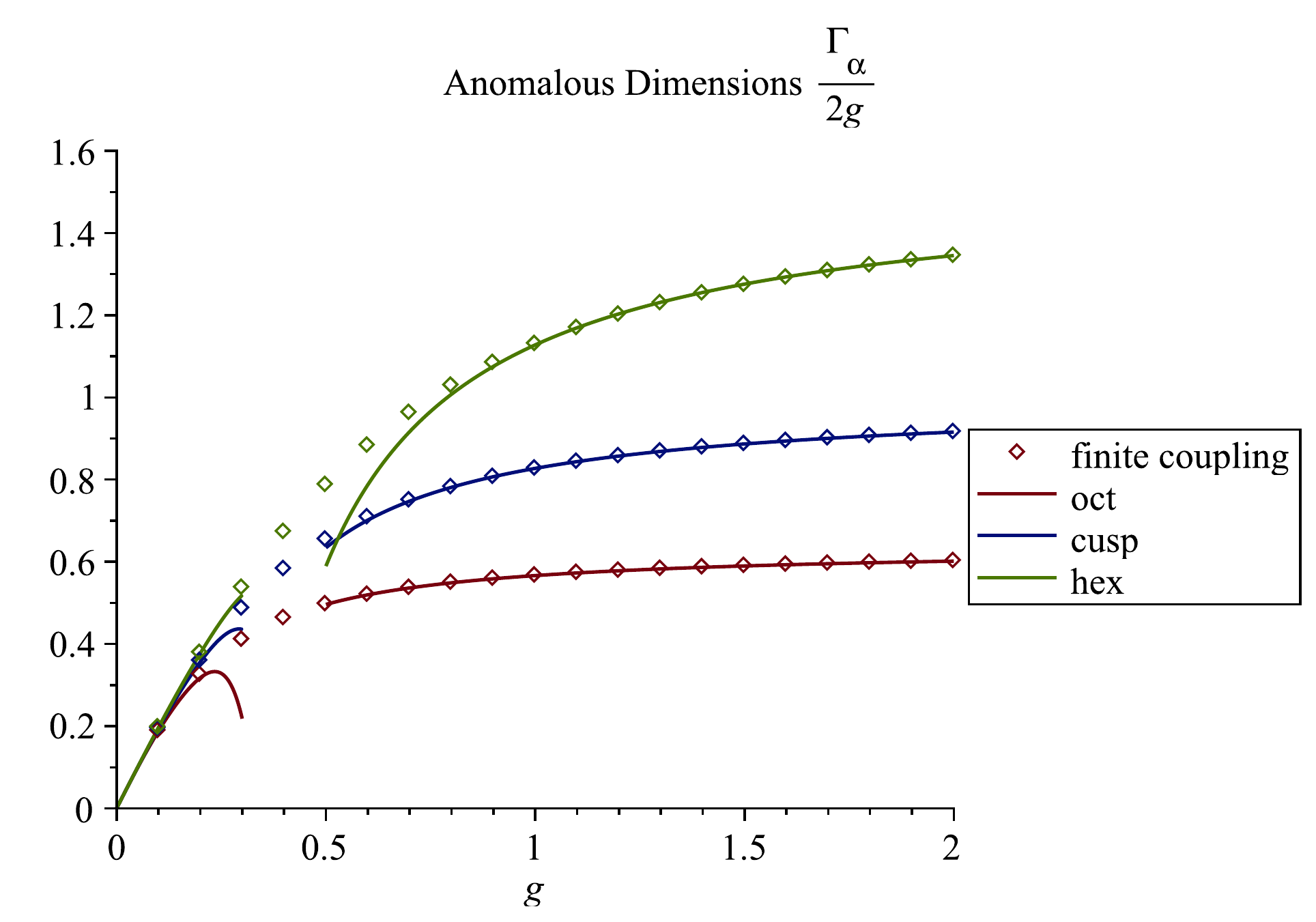}
\end{center}
\vspace{-16pt}
\caption{Plot of $\Gamma_{\alpha}/2g$ as a function of $g$ and comparison with weak and strong coupling expansions, eqs.~(\ref{eq:Ga}) and (\ref{eq:Ga-sc}), respectively.}\label{plot-gam} 
\end{figure}

In Appendix~\ref{appx:strong}, we analyze the strong-coupling behavior, and provide four terms in the expansion of $\Gamma_\alpha$ and
two terms for $D(\alpha)$. Here we quote the leading-order expressions,
\be\label{eq:GalphaDStrong}
\Gamma_{\alpha}\approx \frac{8\alpha g}{\pi\sin{(2\alpha)}}\,,\quad D(\alpha) \approx 4\pi g \Bigl[ \frac{1}{4} - \frac{\alpha^2}{\pi^2} \Bigr]\,.
\ee
In fig.~\ref{plot-gam}, our weak- and strong-coupling expansions are compared with finite-coupling numerics.  The agreement is excellent.

We can also validate our formulae at strong coupling through comparison with string theory, as described in more detail in Appendix~\ref{appx:surface}.
On the diagonal $u = u_{1} = u_{2} = u_{3}$, the string-theoretic analysis yields~\cite{Alday:2009dv,Basso:2014jfa}
\beq\label{eq:R6-strong} 
(\ln{\E(u, u, u)})/\Gc = -\frac{3}{4\pi} \ln^{2}{u} - \frac{\pi^2}{12}
-\frac{\pi}{6} + \frac{\pi}{72}
\eeq
at small $u$, up to power corrections. With the help of \eqref{eq:GalphaDStrong}, we can perfectly reproduce the above result, including the sphere contribution \cite{Basso:2014jfa} of $+\pi/72$. Off the diagonal the behavior is richer at strong coupling. Nonetheless following ref.~\cite{Ito:2018puu} we can also confirm the leading strong coupling behavior of $\Gt=\Gamma_{\pi/3}$ in (\ref{eq:GalphaDStrong}).

\section{Cosmic normalization}

At last, let us remark about the normalization of the amplitude. The subtraction of divergences in the amplitude leaves a freedom in defining the finite part. Depending on the situation, it might prove convenient to subtract more than just the BDS-like amplitude. For example, in the collinear limit it is natural to work with the non-cyclic-invariant object $\W_6$. Another instance is provided by the so-called cosmic normalization for $\mathcal{E}$ introduced in the hexagon function bootstrap,
\beq\label{eq:E-cosmic}
\mathcal{E}_{\textrm{cosmic}} = \mathcal{E} / \rho\, ,
\eeq
with $\rho = \rho(g^2)$ a function of the coupling constant. This function was determined iteratively in \cite{Caron-Huot:2019vjl,Caron-Huot:2019bsq} by demanding that the spaces of functions in which the perturbative amplitudes live obey a coaction principle associated to a cosmic Galois group~\cite{Schnetz:2013hqa,Panzer:2016snt,Brown:2015fyf}. The implementation of this requirement fixes $\rho$ order by order in perturbation theory,
\beq
\begin{aligned}\label{rhoold}
\ln{\rho} = &\,\, 8 \zeta_{3}^2 g^6 -160\zeta_{3}\zeta_{5} g^8 \\
&+ 16(-2 \zeta_{4}\zeta_{3}^2 + 57 \zeta_{5}^2 +105 \zeta_{3}\zeta_{7})g^{10} + \ldots,
\end{aligned}
\eeq
and two more loops can be found in ref.~\cite{Caron-Huot:2019vjl}. Strictly speaking, $\ln\rho$ is fixed up to addition of pure even zeta values, which are trivial under the coaction, and in eq.~(\ref{rhoold}) all pure even zeta values $\zeta(2L)$ have been set to zero.

In the process of evaluating the infinite-dimensional integral in
Appendix~\ref{appx:Gaussian}, a particular normalization factor emerges,
\beq\label{eq:norm}
\N = \textrm{det}\, [\1+\K]\, e^{-\frac{1}{2}\zeta_{2}\Gc}\, ,
\eeq
Remarkably, the perturbative expansion of $\N$ bears a striking resemblance
to $\rho$.  To be precise, one has through at least seven loops
\beq\label{eq:cosmic-conjecture}
\ln\rho - \ln\N\ =\ \textrm{pure even zeta values}.
\eeq
It is tempting to believe that eq.~(\ref{eq:cosmic-conjecture}) holds true to all orders in perturbation theory.  It strongly suggests that the most natural normalization for the amplitude is simply to set $\rho = \N$. This $\rho$ value shifts $C_0$ in eq.~(\ref{eq:C0}) to $C_0 = - D(\pi/3) - \frac{1}{2}D(0)$, removing all $\alpha=\pi/4$ contributions from $\ln \mathcal{E}_{\textrm{cosmic}}$.

%%%%%%%%%%%%%%%%%%%%%%%%%%%%
\section{Conclusion and outlook}

We reported exact expressions for the anomalous dimensions and constant controlling the six-gluon MHV amplitude at the origin of the kinematical space. Our proposals rely on study of the weak coupling series on the field theory side and an extrapolation based on the pentagon OPE formulae. We evaluated our exact expressions to high orders in perturbation theory, numerically at finite coupling, as well as a few orders at strong coupling. The leading strong-coupling behavior was verified to agree with the string theory minimal surface analysis, plus a constant from the sphere determinant.

The main implication of our analysis is that the hexagon amplitude can be determined exactly at the origin, using the same ingredients needed for the cusp anomalous dimension, but tilted by an angle $\alpha$.  For now, the physical significance of $\alpha$ is unclear.  Perhaps similar simplifications and extrapolations will be found for higher polygonal Wilson loops, utilizing other values of $\alpha$.  For example, one can define an ``origin'' of the heptagon by sending six of seven cross ratios to zero; the seventh is not independent and must go to unity.  This limit is currently under investigation.  Weak-coupling expansions generally feature coefficients of zeta values that are rational numbers.  This consideration and eq.~(\ref{eq:G-alpha}) implies that $\sin^2\alpha \in \mathbb{Q}$.
Our work also raises the hope of understanding the behavior at the origin for non-MHV amplitudes, and as one moves away from the origin for both MHV and non-MHV amplitudes, although in both cases it will not be as simple as the quadratic logarithmic behavior explored here.  One could also study light-like Wilson hexagons in other theories, to see whether the integrability of planar ${\cal N}=4$ SYM is critical to this behavior.

We also observed an intriguing connection with the anomalous dimension which controls the light-like limit of the correlator of four half-BPS operators dubbed the octagon \cite{Coronado:2018ypq,Coronado:2018cxj,Belitsky:2019fan}. It is reminiscent of the general correspondence between light-like correlators and null polygonal Wilson loops \cite{Alday:2010zy}. It is not quite the same, however, since the Wilson loop studied here carries no $R$ charge, while the octagon is full of it. It might be hinting at a connection between integrable descriptions based on the polygonalization of correlators \cite{Basso:2015zoa,Fleury:2016ykk,Eden:2016xvg} and amplitudes \cite{Basso:2013vsa}.

%%%%%%%%%%% Acknowledgements
\begin{acknowledgments}
The work of BB was supported by the French National Agency for Research grant ANR-17-CE31-0001-02. The work of LD was supported by the US Department of Energy under contract DE--AC02--76SF00515 and by a Humboldt Research Award. GP acknowledges support from the Deutsche Forschungsgemeinschaft under Germany's Excellence Strategy – EXC 2121 ``Quantum Universe'' – 390833306. BB and LD thank the Pauli Center of ETH Z\"urich and the University of Z\"urich for hospitality. LD is also grateful to the Laboratoire de physique de l'Ecole normale sup\'erieure and Humboldt Universit\"at zu Berlin for hospitality.
\end{acknowledgments}

%%%%%%%%%%%%%%%%%%%%%%%%%%%%%%%%

\appendix

\section{A secretly Gaussian integral}\label{appx:Gaussian}

The key to resumming the OPE at finite coupling close to the origin lies in the structure of its integrand on the Goldstone sheet. Setting $\sigma=\tau=0$ for simplicity (we will recover the full dependence near the origin later, in eq.~(\ref{eq:shift})), it can be written as the product of a Cauchy determinant and a universal Gaussian dressing factor \cite{Basso:2014nra},
\beq\label{eq:factorization}
\frac{\prod_{k}\mu_{k}}{\prod_{k<l}P_{kl}P_{lk}} = \textrm{det}\,\left[ \frac{\hat{\mu}_{k}}{x^{+}_{k}-x^{-}_{l}}\right] \times e^{-\vec{Q} \cdot \G^{-1}\cdot \vec{Q}} \,,
\eeq
with $x^{\pm}_{k} =x^{[\pm a_{k}]}(u_{k}) = x(u_k\pm ia_k/2)$, where $x(u) = \tfrac{1}{2}(u+\sqrt{u^2-4g^2})$ is the Zhukowski variable, and with the reduced measure
\beq
\hat{\mu}_{k} = (-1)^{a_k}\frac{i x^{+}_{k} x^{-}_{k}}{\sqrt{((x^{+}_{k})^2-g^2)((x^{-}_{k})^2-g^2)}}\, .
\eeq
Here $\vec{Q} = \sum_{k=1}^{N}\vec{q}\, (u_{k}, a_{k})$, with $\vec{q} = (q_{j=1, 2, \ldots }^{\pm})$, is a vector of higher conserved charges with components~\footnote{These charges are related to the coefficients $k_{j}, \tilde{k}_{j}$ defined in Appendix B of \cite{Basso:2014nra} by $q_{j}^{\pm} = k_{j}\pm i \tilde{k}_{j}$.}
\beq\label{eq:charges}
q^{\pm}_{j}(u, a) = \frac{(ig)^{j}}{2j} \sum_{l=-\frac{1}{2}(a-1)}^{\frac{1}{2}(a-1)} \left[\frac{1}{(x^{[\pm 1-2l]})^{j}}\pm \frac{1}{(-x^{[\mp 1-2l]})^{j}} \right]\, .
\eeq
They are contracted with the inverse of the symmetric form $\G$ which acts trivially on the upper indices $m, n \in \{+, -\}$,
\beq
\G^{mn} = \delta^{mn}\,  (\1+\K)\cdot \Q^{-1}\, ,
\eeq
and as the kernel $\K$ of the BES equation on the lower indices, see eq.~(\ref{eq:K}). Here, $\Q$ is a diagonal semi-infinite matrix, with elements $\Q_{ij} = j (-1)^{j+1} \delta_{ij}$.

As a result of this factorization, the gluonic contributions can be written concisely as an infinite-dimensional integral,
\beq\label{eq:FG}
\E = \N \int  \prod_{i=1}^{\infty}d\xi_{i}^{+}d\xi_{i}^{-}  \F_{\varphi}(\vec{\xi} \, )\, e^{-\vec{\xi}\cdot \G \cdot \vec{\xi}}\, ,
\eeq
where $\vec{\xi}$ is a vector of variables conjugate to the charges (\ref{eq:charges}) and where $\N$ is the normalization factor,
\beq\label{eq:normappendix}
\N = \textrm{det}\, [\1+\K]\, e^{-\frac{1}{2}\zeta_{2}\Gc}\, ,
\eeq
up to an irrelevant coupling independent factor. The integrand $\F_{\varphi}$ is a Fredholm determinant which generates the prefactor in (\ref{eq:factorization}),
\beq\label{eq:CFdet}
\begin{aligned}
&\ln{\F_{\varphi}} = -\sum_{N\geqslant 1}\frac{1}{N}\sum_{\textbf{a}}\varointctrclockwise \frac{d\textbf{u}}{(2\pi)^{N}} \prod_{k=1}^{N}\frac{\hat{\mu}_k e^{\varphi a_{k}}}{x_{k}^{+}-x_{k+1}^{-}} \, e^{2i\vec{Q}\cdot \vec{\xi}}\, ,
\end{aligned}
\eeq
with $x_{N+1}^{-} \equiv x_{1}^{-}$ and with the contour along $|x^{-}_{k}| = g$. This ingredient appears similar to the functional determinant representing the octagon correlator \cite{Kostov:2019stn}. It differs in that it is not only a function of the cross ratios, but also of the infinite set of dummy variables $\vec{\xi}$ \footnote{Similar objects -- so-called $\tau$-functions -- also appear naturally in the study of matrix-model integrals.}.

The dependence on the variables $\sigma$ and $\tau$ is recovered by adding energy and momentum factors inside the integrals in (\ref{eq:CFdet}), using known formulae for the total energy and momentum of an $N$-gluon state on the Goldstone sheet~\cite{Basso:2014nra}
\beq
E \mp i P =  (1\pm 1)\sum_{k=1}^{N} a_{k} + 4g [M^{-1}\cdot \vec{Q} \, ]_{1}^{\pm}\, .
\eeq
In addition, when considering $\E$, one must include the $\tau$ and $\sigma$ dependent part of the overall framing factor in (\ref{eq:framedW}), controlled by the cusp anomalous dimension $4g^2 [M^{-1}]_{11}^{\alpha\beta} = \Gc \delta^{\alpha\beta}$. The two effects can be obtained at once starting from the $\sigma = \tau = 0$ integral (\ref{eq:FG}) by letting $\varphi\rightarrow \varphi-\sigma-\tau$ in (\ref{eq:CFdet}) and including the exponential factor
\beq\label{eq:shift}
\exp{2ig (\tau+\sigma) \xi^{+}_{1}+ 2ig (\tau-\sigma) \xi^{-}_{1}}
\eeq
inside the integrand in (\ref{eq:FG}). This prescription is readily verified, after shifting the integration variables,
\beq
\xi^{\pm}_{j} \rightarrow \xi_{j}^{\pm} +i g (\tau \pm \sigma) [M^{-1} ]^{\pm\pm}_{j1} \, ,
\eeq
and using the aforementioned formulae for $E, P$ and $\Gc$.

Representation (\ref{eq:FG}) can be evaluated at weak coupling after noticing that the charges $Q_{i} \sim g^i$. One can thus Taylor expand the exponential in (\ref{eq:CFdet}) around $\vec{\xi} = \vec{0}$,
\beq\label{eq:cumulant}
\ln \F_{\varphi}(\vec{\xi}\, ) = \mathcal{h}1\mathcal{i} + 2i\mathcal{h}Q^{m}_i\mathcal{i} \xi_{i}^{m} -2 \mathcal{h}Q^{m}_i Q^{n}_j\mathcal{i} \xi_{i}^{m} \xi_{j}^{n} +\ldots\, ,
\eeq
with implicit sums over lower and upper indices; the coefficient at order $L$ is $\sim g^{L}$ or smaller. Extra simplification comes from the structure of the rapidity integrals, which cause the sum over $N$ to truncate at $N = L$ at $L$ loops.

Quite remarkably, the series (\ref{eq:cumulant}) is observed to truncate at large $\varphi \rightarrow \infty$. Namely, generating expressions to higher loops, we observed that the expansion in $\vec{\xi}$ terminates at quadratic order, or, equivalently, that \textit{all} moments of degree $> 2$ vanish at large $\varphi$,
\beq
\label{threeormorevanish}
\lim_{\varphi\rightarrow \infty}\mathcal{h}Q_{i}^{m} Q_{j}^{n}\, Q_{k}^{p} \ldots  \mathcal{i} = 0\, .
\eeq
The non-zero $k$-moments are found to be of degree $2-k$ in $\varphi$. This truncation immediately implies the double logarithmic behavior of $\ln{\E}$ at the origin.

%\section{Conjectures for the moments}\label{appx:Moments} - merged!!!!

Furthermore, investigation of the non-zero moments led us to simple all-order conjectures for the moments entering the secretly Gaussian integral (\ref{eq:FG}). From the observed Gaussian behavior~(\ref{threeormorevanish}), the logarithm of the Wilson loop at the origin is characterized by a quadratic form and a vacuum expectation value $\langle\vec{Q}\rangle$,
\beq\label{eq:Gauss}
\ln{\E} = - \mathcal{h}\vec{Q}\mathcal{i} \cdot \frac{1}{\G+ \delta\G}\cdot\mathcal{h}\vec{Q}\mathcal{i} + V \, ,
\eeq
where $\delta\G^{mn}_{ij} = \lim_{\varphi\rightarrow \infty}2\mathcal{h}Q_{i}^{m}Q_{j}^{n}\mathcal{i}$ is the shift of the quadratic form, and with
\beq
V = \mathcal{h}1\mathcal{i}-\frac{\zeta_{2}}{2}\Gc + \frac{1}{2}\ln{\textrm{det}\left[\frac{\G}{\G+\delta \G}\right]} \, .
\eeq
Equation~(\ref{eq:Gauss}) is applicable to $\sigma = \tau = 0$, which gives us access to a linear combination of $\Go$ and $\Gt$. The general case requires taking into account the small modifications described around eq.~(\ref{eq:shift}).

Now, we observed empirically, through four loops, that all moments can be expressed in terms of the building blocks of the BES kernel given in eq.~(\ref{eq:blocks}). More precisely, we found that $\delta\G$ is diagonal in the upper indices, with
\beq\label{eq:Gpp}
\delta\G^{++}\cdot \mathbb{Q} = \frac{1}{2}\left[\begin{array}{cc} \K_{\circ\circ}\frac{1}{\1+\K_{\circ\circ}} &-\frac{1}{\1+\K_{\circ\circ}}\K_{\circ\bullet} \\ -\K_{\bullet\circ}\frac{1}{\1+\K_{\circ\circ}} &  -\K_{\bullet\bullet}- \K_{\bullet\circ}\frac{1}{\1+\K_{\circ\circ}}\K_{\circ\bullet} \end{array}\right]\, ,
\eeq
and similarly for $\delta\G^{--}$ after permuting lines, columns and subscripts $\circ\leftrightarrow \bullet$. We also observed that
\beq
\begin{aligned}\label{Qand1}
\mathcal{h}Q^{+}_i\mathcal{i} &= \frac{g\varphi}{2} (\delta_{i1} - 2 \delta\G^{++}_{i1})\, ,  \qquad \mathcal{h}Q^{-}_i\mathcal{i} = -\frac{g\varphi}{2} \delta_{i1}\, , \\
\mathcal{h}1\mathcal{i} &= - \frac{g^2 \varphi^2}{2} (\1+\K_{\circ\circ})^{-1}_{11} - \frac{1}{2}(D_{\circ\circ}+D_{\bullet\bullet})\, ,
\end{aligned}
\eeq
with $D_{\circ\circ} = \ln \textrm{det}\,  [1+\K_{\circ\circ}]$ and similarly for $D_{\bullet\bullet}$.

As a cross check of our conjectures, we verified, after reinstating the three cross ratios using (\ref{eq:shift}), that the final prediction for $\E$ is permutation symmetric and can be cast into the form (\ref{eq:main}). This step requires some elementary algebra for block matrices, see e.g.~ref.~\cite{BlockMatrix}. Lastly, similar algebra can be used to simplify the expressions and derive the concise formulae (\ref{eq:G-alpha}) and (\ref{eq:C0}).

%%%%%%%%%%%%%%%%%%%%%%
\section{Strong coupling analysis}\label{appx:strong}

In this appendix we examine the strong coupling regime $\sqrt{\lambda} = 4\pi g \rightarrow \infty$. This regime is harder to address than weak coupling because the rank of the matrix $\K(\alpha)$ scales like $g$, and thus the matrix truly is infinite dimensional at large $g$. Nonetheless, the problem can be solved by going to an alternative representation \cite{Alday:2007qf,Basso:2007wd,Kostov:2008ax,Basso:2009gh}. Define the infinite vector
\beq
\vec{v}(t) = [iJ_{1}(t), -J_{2}(t), iJ_{3}(t), -J_{4}(t), \ldots]\,.
\eeq
(Note that it has no upper index, unlike the vectors introduced in the main text.)
Then the inversion problem is equivalent to calculating the function
\beq
\gamma(t, s) = \gamma(s, t) = -\vec{v}(t) \cdot \mathbb{Q}\cdot [1+\K(\alpha)]^{-1} \cdot \vec{v}(s)\, .
\eeq
The latter is an entire function in both $s$ and $t$, with Fourier transform in each variable supported on the interval $(-1, 1)$. One then notices that the problem can be cast into the form of a Riemann-Hilbert equation,
\beq\label{eq:RH}
\int\limits_{-\infty}^{\infty}dt\, e^{iut}\Omega(t, s) \left(\cos{\alpha} + i \sin{\alpha}\, \textrm{sgn}\, {t}\right)  = e^{ius}\, ,
\eeq
with $u \in (-1, 1)$ and where
\beq\label{eq:transf}
\Omega(t, s) =  \frac{\cosh{(\tfrac{t}{4g}-i\alpha)}}{s\sinh{(\tfrac{t}{4g})}} \gamma(t, s)\, .
\eeq
The nice thing about this formulation is that the coupling constant $g$ only enters in the transformation (\ref{eq:transf}). It permits us to solve the problem by first obtaining a general solution for $\Omega$ and then implementing the analyticity requirements on the solution. 

Once the solution is known, one reads off the anomalous dimension using
\beq\label{eq:G-str}
\Gamma_{\alpha} = 16g^2 \lim_{s, t\rightarrow 0} \frac{\gamma(t, s)}{s t} =\frac{4g\, \Omega(0, 0)}{\cos{\alpha}} \, ,
\eeq
whereas, for computing the determinant $D(\alpha)$, one can rely on
\beq
\partial_{\alpha}D(\alpha) = \textrm{tr}\, \left[\frac{\partial_{\alpha}\K(\alpha)}{1+\K(\alpha)}\right] =  2\Re\, \int\limits_{0}^{\infty}dt \frac{ie^{2i\alpha-\frac{t}{4g}}\Omega(t, t)}{\cosh{(\frac{t}{4g}-i\alpha)}}\,,
\eeq
and the exact relation (\ref{eq:exactD}) for the constant of integration $D(0)$.

\begin{figure}
\begin{center}
\includegraphics[scale=0.45]{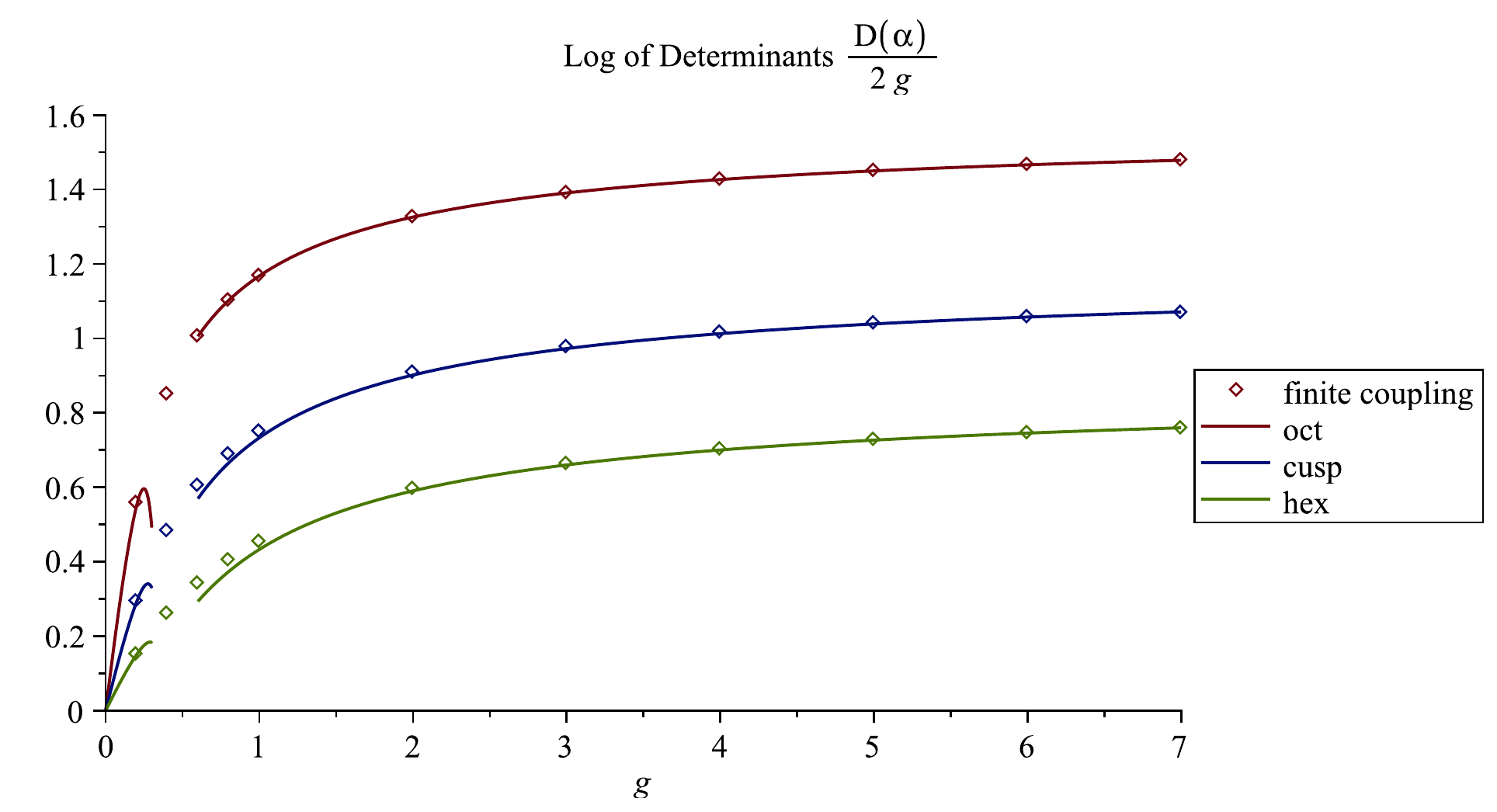}
\end{center}
\caption{Plot of $D(\alpha)/2g$ as a function of $g$. The solid lines show the weak and strong coupling estimates, obtained using (\ref{eq:Da}) and (\ref{eq:D-str}).}\label{plot-det} 
\end{figure}

A general method for solving this type of problem was proposed in \cite{Basso:2007wd,Kostov:2008ax,Basso:2009gh} for the case $\alpha = \pi/4$ and for $s=0$. It extends smoothly to a generic value of $\alpha$ and for $s\neq 0$. To leading order at strong coupling, the solution is given by a particular solution to (\ref{eq:RH}) with Fourier transform supported on the interval $(-1, 1)$,
\beq\label{eq:part}
\Omega(it, is) = \frac{t V_{0}(t) V_{1}(s)-s V_{0}(s) V_{1}(t)}{V_{1}(0)(t-s)} +\ldots  \,,
\eeq
where the dots stand for terms that are subleading at large $g$, for $t, s = O(1)$, and where $V_{0, 1}$ are special functions,
\beq
\begin{aligned}
V_{r}(t) = \int\limits_{-1}^{1}\frac{du}{2\pi}\, (1+u)^{\alpha/\pi-r}(1-u)^{-\alpha/\pi} e^{u t} \, .
\end{aligned}
\eeq
The latter can also be written in terms of hypergeometric functions, for $r=0,1$,
\beq
V_{r}(t) = \frac{(2\alpha/\pi)^{1-r}}{2\sin{\alpha}} e^{-t} \,_{1}F_{1}(\alpha/\pi+1-r, 2-r, 2t)\,.
\eeq
Plugging solution (\ref{eq:part}) inside (\ref{eq:G-str}) we get
\beq\label{eq:Ga-sc-lead}
\Gamma_{\alpha} = \frac{8\alpha g}{\pi\sin{(2\alpha)}} + O(g^0)\, ,
\eeq
whereas for the determinant it yields,
\beq\label{eq:D-str}
D(\alpha) = 4\pi g \Bigl[ \frac{1}{4} - \frac{\alpha^2}{\pi^2} \Bigr]
  - \Bigl[ \frac{1}{4}+\frac{\alpha^2}{\pi^2} \Bigr] \ln{(4g)} + C(\alpha) + \ldots\, .
\eeq
The numerical agreement between the strong coupling expansions for $D(\alpha)$ and a finite-coupling evaluation is excellent, as shown in fig.~\ref{plot-det}. The constant $C(\alpha)$ is not determined by the particular solution alone and receives corrections from subleading terms in (\ref{eq:part}). We fitted it for $\alpha\neq0$ to the values
\beq\label{Cfit}
C(\pi/4) =  -0.457\, , \qquad C(\pi/3) =  -0.379\, ,
\eeq
which are close to the exact value for $\alpha = 0$, given by $C(0) = -\frac{1}{4}\ln{(2\pi)} = -0.459$.

The subleading terms in (\ref{eq:part}) are obtained by adding a homogeneous solution to the Riemann-Hilbert equation of the form \cite{Basso:2009gh}
\beq
\delta \Omega(it, is) = f_{0}(t, s) V_{0}(t) + f_{1}(t, s) V_{1}(t)\, ,
\eeq
where $f_{0, 1}$ are two meromorphic functions of $t$ with simple poles at $t = 4\pi m g$ with $m\in \mathbb{Z}_{\neq 0}$. The latter functions are determined by their asymptotics at large $t$ and the requirement that the full solution, which is the sum of the particular and the homogeneous solution, has zeros at $t = 4g(\alpha-\pi (m-\tfrac{1}{2}))$ with $m\in \mathbb{Z}$, due to the numerator on the right-hand side of eq.~(\ref{eq:transf}). The algorithm is explained in great detail in \cite{Basso:2009gh} for $\alpha = \pi/4$ and $s=0$. It works the same for generic $\alpha$ and $s$. For the determination of $\Gamma_{\alpha}$, one can specialize to $s=0$.
Skipping the intermediate steps, we simply quote here the end result for the first few terms in the expansion of $\Gamma_{\alpha}$. They read
\beq\label{eq:Ga-sc}
\Gamma_{\alpha} = \frac{8ag}{\sin{(2\pi a)}}\left[1 - \frac{s_{1}}{2\sqrt{\lambda}} - \frac{a s_{2}}{4\lambda} - \frac{a(s_{1}s_{2}+as_{3})}{8(\sqrt{\lambda})^3}+\ldots\right],
\eeq
with $a=\alpha/\pi$ and where $s_{k}$ are the coefficients in
\beq\label{eq:S}
\frac{\Gamma(\tfrac{1}{2}+a)\Gamma(\tfrac{1}{2}-a+t)\Gamma(1-t)}{\Gamma(\tfrac{1}{2}-a)\Gamma(\tfrac{1}{2}+a-t)\Gamma(1+t)} = \exp \sum_{k=1}^{\infty} \frac{s_{k}(-t)^k}{k!}\, ,
\eeq
that is,
\beq
s_{k+1} = \{\psi_{k}(1)-\psi_{k}(\tfrac{1}{2}+a)\} + (-1)^{k} \{\psi_{k}(1)-\psi_{k}(\tfrac{1}{2}-a)\}\, ,
\eeq
with $\psi_{k}(z) = \partial_{z}^{k+1}\ln{\Gamma(z)}$.

These formulae generalize to generic $a$ the ones obtained for $\Gc$. When $a=0$ the series truncates at one loop in agreement with the exact representation (\ref{eq:exact}). For $a=1/3$ one obtains the strong-coupling expansion of the new anomalous dimension $\Gt$.

We should add that in general, like for $\Gc$, the strong coupling series appears divergent and non-Borel summable \cite{Basso:2007wd}. It signals the need to include an additional expansion parameter $\Lambda^2$, which is exponentially small at strong coupling, for fully determining the solution.  (The resurgence property of this transseries was discussed in \cite{Aniceto:2015rua,Dorigoni:2015dha} for the cusp anomalous dimension.) Following the analysis in \cite{Basso:2009gh}, we have found
\beq
\Lambda^2 \sim \lambda^{a} e^{-(1-2a)\sqrt{\lambda}}
\eeq
for $0\leqslant a < 1/2$. For $a=0$ it agrees with the size of the exponentially small corrections in (\ref{eq:exact}). When $a = 1/4$, $\Lambda$ was given a physical meaning and associated to the mass gap of the O(6) sigma model \cite{Alday:2007mf,Basso:2009gh}, which enters as the low-energy effective theory for the flux tube. Its physical significance for other values of $a$ is mysterious.

\section{Minimal surface analysis}\label{appx:surface}

Our findings can be compared with the string theory analysis at strong coupling. According to the holographic dictionary, the vacuum expectation value of the Wilson loop is given by the open-string path integral for a string ending on the polygonal contour at the boundary of Anti-de-Sitter space (AdS) \cite{Maldacena:1998im,Rey:1998ik}. The latter can be evaluated semi-classically at strong coupling,
\beq\label{lnEstrong}
(\ln{\E})/\Gc = -A_{6} -\frac{1}{2}(\sigma^2+\tau^2+\zeta_{2})+ \frac{\pi}{72}\, ,
\eeq
with $\Gc \approx 2g$ the string tension \cite{Gubser:2002tv}.  Here the first term is minus the (renormalized) area of a minimal surface in AdS ending on the polygonal contour of the loop at the boundary of AdS \cite{Alday:2007hr,Alday:2009dv}. It is given by the Yang-Yang functional of an associated system of thermodynamic Bethe ansatz (TBA) equations \cite{Alday:2009dv,Alday:2010ku}. The middle term results from the definition of $\E$ and the last term \cite{Basso:2014jfa} is a shift coming from the determinant of the quantum fluctuations along the 5-sphere.

Our predictions are easily checked along the diagonal $u = u_{1} = u_{2} = u_{3}$. The TBA equations are exactly solvable in this case, and yield~\cite{Alday:2009dv}
\beq\label{eq:R6-strong-app} 
(\ln{\E})/\Gc = -\frac{3}{4\pi} \ln^{2}{u} - \frac{\pi^2}{12}-\frac{11\pi}{72}
\eeq
at small $u$, up to power corrections. The coefficient of $\ln^{2}{u}$ agrees perfectly with (\ref{eq:main}), using the $g\to\infty$ limit of the exact formula (\ref{eq:exact}) for $\Go$: $\Go \approx (2/\pi) \times \Gc$. The comparison for the constant requires eq.~(\ref{eq:D-str}) for the determinants entering $C_0$, and it also works analytically. 

The third anomalous dimension $\Gt$ is associated to off-diagonal behavior. This regime is harder to probe, as the TBA equations can no longer be solved exactly. Nonetheless, following ref.~\cite{Ito:2018puu}, we find that the equations simplify when $\varphi, \tau \rightarrow \infty$, keeping their ratio $\varphi/\tau$ fixed.  Namely, they can be cast as a single linear integral equation describing a condensate of positive-helicity gluons. Setting $\sigma = 0$ for simplicity and using the TBA equations in the form given in Appendix F of ref.~\cite{Alday:2010ku}, one finds
\beq\label{eq:area}
A_{6} \cong \int\limits_{-B}^{B}\frac{d\theta}{2\pi} f(\theta)I(\theta) +\frac{\pi}{6}\, ,
\eeq
where $f(\theta)$ solves the equation
\beq
f(\theta) = I(\theta) + \int\limits_{-B}^{B}\frac{d\theta'}{2\pi} K(\theta-\theta') f(\theta')\, ,
\eeq
with $K = \sech{\theta}$ and $I = (\varphi-\sqrt{2}\tau \cosh{\theta})\, \sech{(2\theta)}$. Here $f(\theta)$ describes the rapidity distribution of gluons with mass-to-charge ratio $\sqrt{2}$. This function is positive on the support $(-B, B)$ and the Fermi rapidity $B$ is determined self-consistently by demanding that $f(\pm B) = 0$. Furthermore, $\varphi/\tau \geqslant  \sqrt{2}$ for a real solution to exist.

The near-diagonal limit corresponds to letting $B\rightarrow \infty$.  Setting $B=\infty$, the solution is found immediately by going to a Fourier space, $\hat{f}(s) \equiv \int_{-\infty}^{\infty} \frac{d\theta}{2\pi} f(\theta) \cos{(s\theta)}$:
\beq\label{eq:fofs}
\hat{f}(s) = \hat{I}(s)/[1-\hat{K}(s)]\, ,
\eeq
with $\hat{I}(s) = \tfrac{1}{4}\sech{(\tfrac{\pi s}{4})} \left[\varphi-\tau - \tau\sech{(\tfrac{\pi s}{2})}\right]$ and $\hat{K}(s) = \tfrac{1}{2} \sech{(\tfrac{\pi s}{2})}$ the Fourier transforms of the source term and kernel, respectively. Plugging $\hat{f}(s)$ into (\ref{lnEstrong}) and (\ref{eq:area}) yields,
for $\sigma = 0$,
\beq\label{eq:Efin}
\ln{\E} = -\frac{\Go}{6} \varphi^2-\frac{\Gt}{12}(\varphi-3 \tau)^2 + C_0 \,,
\eeq
with the strong coupling values
\beq
\frac{\Go}{\Gc} = \frac{2}{\pi}\, , \qquad \frac{\Gt}{\Gc} = \frac{8}{3\sqrt{3}}\, ,
\eeq
in perfect agreement with the flux-tube prediction~(\ref{eq:Ga-sc-lead}).

Lastly, we should stress that this matching comes with a caveat.
It traces back to the fact that the assumption that $B = \infty$ is valid on the diagonal ($\varphi = 3\tau$) but not away from it. This is verified by mapping (\ref{eq:fofs}) to $\theta$-space and noticing that $f(\theta)$ turns negative at $B \sim \frac{3}{4}\ln{(\frac{2\varphi}{3\tau-\varphi})}$, which is large for $3\tau-\varphi \sim 0$ but not infinite. As a result, the strong-coupling formula (\ref{eq:Efin}) only holds up to corrections arising from the finiteness of $B$. For consistency, it must be that the finite-$B$ corrections correspond to terms that we discarded in the finite-coupling analysis, because they were power suppressed in the cross ratios (at finite coupling). On the other hand, the finite-$B$ corrections to the minimal surface area cannot be power suppressed, since they must take the form $\varphi^2 F(\frac{3\tau-\varphi}{\varphi})$ for some $F$. This paradox hints at an order of limits issue, between the strong-coupling limit and the approach to the origin, away from the diagonal. It would be very interesting to study this phenomenon in more detail and determine the form of the function $F$, using more powerful techniques like the one developed in \cite{Ito:2018puu}.

% Bibliography

%merlin.mbs apsrev4-1.bst 2010-07-25 4.21a (PWD, AO, DPC) hacked
%Control: key (0)
%Control: author (8) initials jnrlst
%Control: editor formatted (1) identically to author
%Control: production of article title (-1) disabled
%Control: page (0) single
%Control: year (1) truncated
%Control: production of eprint (0) enabled
%

\end{document}